\documentstyle{article}
\begin{document}

\title
{Interference and entanglement: an intrinsic approach}

\author{\large   Vladimir I  Man'ko,$^1$
Giuseppe Marmo,$^2$ E C George Sudarshan$^3$
and Francesco Zaccaria$^2$      \\         
%
\normalsize
${}^1$P. N. Lebedev Physical Institute, Leninskii Prospect 53,
Moscow 119991 Russia\\
\normalsize
${}^2$Dipartimento di Scienze Fisiche,
Universit\`a ``Federico II'' di Napoli  \\
\normalsize
and Istituto Nazionale di
Fisica Nucleare, Sezione di Napoli, \\
\normalsize
 Complesso Universitario di
Monte S.~Angelo, Via Cintia, I-80126 Napoli, Italy\\
\normalsize
${}^3$Physics Department, Center for Particle Physics, University of Texas,
78712 Austin, Texas   \\            
\normalsize
E-mail:  \\
\normalsize
manko@sci.lebedev.ru  \\
\normalsize
marmo@na.infn.it  \\
\normalsize
sudarshan@physics.utexas.edu  \\
\normalsize
zaccaria@na.infn.it  }

\date{5 July 2002}
\maketitle


\begin{abstract}
An addition rule of impure density operators, which provides a pure
state density operator, is formulated. Quantum interference
including visibility property is discussed in the context of the density
operator formalism. A measure of entanglement is then introduced as the
norm of the matrix equal to the difference between a bipartite density 
matrix and the tensor product of partial traces.
Entanglement for arbitrary quantum observables for multipartite
systems is discussed.
Star-product kernels are used to map the formulation of the
addition rule of density operators onto the addition rule 
of symbols of the operators. Entanglement and nonlocalization of the
pure state projector and allied operators are discussed.
Tomographic and Weyl symbols (tomograms and Wigner functions) are
considered as examples. The squeezed-states and some spin-states
(two qubits)
are studied to illustrate the formalism.
\end{abstract}

{\bf keywords}: purification, entanglement, star-product, superposition


\section{Introduction}

Superposition principle of quantum states plays a key role in such
physical phenomena as interference of matter waves~\cite{Dirac}.
Wave properties of electron are connected with de Broglie wave
length expressed in terms of particle momentum~\cite{DeBroglie}.
These properties are naturally described by a wave function
associated with the particle's quantum state and obeying
Schr\"odinger equation~\cite{Schrodinger26}. For a system with
several degrees of freedom, the possibility to consider two
subsystems --- first one connected with some of degrees of freedom
and the second one with the rest degrees of freedom, respectively,
the superposition principle provides a construction of entangled
states~\cite{Schrodingers}. Discussing two subsystems of a given
system implies the physical possibility to measure characteristic
properties distinguishing the subsystems. 

Entangled states are the
states which are constructed as a superposition of states, each of
which has the wavefunction expressed as a product of wavefunctions
depending on the different degrees of freedom. The mixed states of
quantum systems are described by density operator~\cite{vonNeumann}. 
The superposition principle of pure quantum states has been formulated 
in [6--8] 
in terms of a new addition rule of the density operators. This
addition rule corresponds to a purification procedure of a mixed
quantum state obtained by the standard addition rule of the
density operators. Relation of the purification procedure to
reconstructing the entanglement structure of the mixed state of a
bipartite system has been preliminarily discussed in \cite{Sud3}.

Various notions of measure of entanglement were suggested in [9--13].
All these measures are related to
some operators associated to a bipartite quantum system. The aim
of our work is to give the new addition rule of density operators
describing the superposition of impure density matrices and this 
analysis generalizes the results of [6--8]
where the coherent addition rule of pure density operators was formulated. 
We also define the measure of
entanglement of bipartite and multipartite quantum systems
considering intrinsic properties of the density operator
describing the state. 

Since the density operators can be
considered using different representations for their symbols like,
for example, Wigner function~\cite{Wigner32}, Husimi--Kano
function~\cite{Husimi40,Kano} as well as singular
quasidistribution~\cite{Sudarsh,Glauber}, we discuss the addition
rule of the density operators in terms of the addition rule of
their symbols. To do this, we discuss the star-product of the
operator symbols (see, for example, [19--23]).
We consider also the case of density operator representation by the
standard tomographic probability distribution, which is used to
give a ``probability'' formulation of quantum
mechanics~\cite{ManciniPLA}. We also give the formulation of both
aspects of superposition principle, namely, the coherent addition 
of impure density operators and measure of entanglement in terms of the
star-product quantization procedure.

The paper is organized as follows.

In section~2 the basic ideas of the construction of purification
procedure and measure of entanglement are described. In section~3 the
review of addition rule of pure density operators is presented
including the presence of visibility parameter. In section~4 a new
formula for purification of sum of impure density operators is
obtained. In section~5 a short review of the star-product
formalism is given. The purification formula for a symbol for an
arbitrary kind of density operator which is obtained by
purification of sum of the symbols of impure density operators is
obtained in section~6. The example of the Wigner function addition
rule in terms of star-product kernel is presented in section~7.
Addition rule of tomographic symbols is considered in section~8. 
In section~9 the notions of entanglement and measure of entanglement are
discussed in terms of intrinsic properties of the density operator of
a composite system, while in section~10 a notion of entanglement is
introduced for other observables. The example of two
qubits is considered in section~11. A measure of entanglement of 
multimode squeezed state is presented in section~12.
Purification procedure for separable density matrix is discussed in
section~13. In section~14 the role of fiducial projector used to
formulate the superposition principle in terms of density operators
 is considered.
In conclusions (section~15) some perspectives are discussed, while in
Appendix we give the proof of the theorem that for pure state
$\rho_{AB}$ of bipartite system $AB$ eigenvalues and ranks of
reduced density operators $\rho_A$ and $\rho_B$ are equal.

\section{General ideas}

The notion of dynamical variables generating an algebra of
observables and the notion of states, which are dual to this
algebra, are common to both classical dynamics and quantum
dynamics; but in quantum dynamics the operators generating 
transformations form a noncommutative vector space while the classical
algebra of dynamical variables is commutative. 
Since every true representation
(realization) of a commutative algebra is one-dimensional, this is
no longer true of a noncommutative algebra. This has the immediate
consequence that while one can have states in which all dynamical
variables have unique values in classical dynamics, this is not so
in quantum dynamics. We have fundamental states in quantum
dynamics in which all but a (commuting) set of variables have
definite (dispersion-free) values for which other dynamical
variables have a distribution of values. In the case where the
eigenstates are continuous, this finds expression in generalized
uncertainty relations [25--28].
The generalised uncertainty relations exist also for discrete 
observables like spin (see, e.g., \cite{Vol183}). 

Classical states may be associated with distributions in phase
space (that is, they have distributions in values for all dynamical
variables) of which the extremal (pure) states are points in phase
space (their distributions correspond to Dirac delta functions or
Kronecker delta matrices for the fundamental dynamical variables).
In contrast, the quantum states are linear functionals on the
dynamical variables which map nonnegative operators into
nonnegative numbers. A linear distribution that can be expressed
as a convex combination (sum or integral) of other appropriate
ones can be made up of extremal linear functionals which cannot be
so decomposed. These extremal states are the `pure states' of a
quantum system.

If all quantum states had distributions for some set of dynamical
variables, how can we combine them to get dispersion free states
for these dynamical variables? This depends on the characteristic
quantum property of superposition of states. Thus, for example, in
interference we compose two pure states to form a new pure state.
In terms of states considered as linear functionals, the natural
process is to generate convex linear sums which are not pure
states. But processes like interference, diffraction and 
composition of light polarizations need a new procedure.

An alternate formalism suitable for incorporating this composition
law is given by the vector (Hilbert) space formalism --- with each
pure state we associate a vector (and its dual vector), and the
expectation values are obtained by the dynamical variables acting
on the state as linear operators, and forming the scalar product
with the original dual. Real classical variables have their
counterpart in selfadjoint linear operators and their expectation
value in any state is real. Moreover, positive operators
(nonnegative operators) have nonnegative expectation values. Since
vectors permit linear combinations (over the field of complex
numbers), superposition of states and interference are naturally
explained. Since the sesquilinear tensor product of the vector and
its dual is a linear functional, this can be compared with the
formalism in terms of linear functionals. But the only linear
functionals obtained as a sesquilinear form (outer product of a
vector and its dual) are extremal. To get full correspondence with
the linear functional formulation, we should form convex
combinations of such outer products. Density matrices (nonextremal 
states) are convex combinations of bilinears in the vectors.

Since the sesquilinear form (outer product) of states constitute
pure states, we see that every `state vector' corresponds to an
extremal linear functional, which may be treated as a linear
operator in the vector space, we should identify mixed states as
nonextremal states of the convex set of states. 
But the correspondence of a state vector and its pure density
 operator is not one-to-one. The von Neumann `measurement' is a 
 projector to an eigenstate of particular self-adjoint dynamical
 variable. (In the case spectrum is continuous, von Neumann 
 prescribes a nested family of spectral projections.) The prescription
 may be viewed as imparting a measure on the spectrum of the operator
 representing the dynamical variable.
 
 A generalization of this protocol is called POVM (positive 
 operator valued measure). A von Neumann measurement results 
 in a pure state of the system (or more generally a density matrix
 in the eigenspace of the operator). But a POVM may result in a 
 mixed density matrix. The entire set of vectors $\{\psi e^{i\theta}\}$
  corresponds to a unique extremal state $$\rho=\psi\psi^\dagger.$$ 
  This extremal
operator is idempotent, selfadjoint (of trace class) and satisfies
$$\rho^2=\psi\psi^\dagger\psi\psi^\dagger=\psi\psi^\dagger=\rho$$
since $\psi^\dagger\psi=1$ for normalization. This equivalence
class $\{e^{i\theta}\psi\}$ is identified as a ray --- so extremal
density matrices correspond one-to-one to rays in the vector
(Hilbert) space. Rays do not constitute a vector space. Considered
as a linear operator, since $\rho^2=\rho$ and
Tr$\,\rho=\psi^\dagger\psi=1$, $\rho$ is a projector of rank one.
So extremal states are projectors associated with unique rays. A
mixed state is a probabilitistic (nonnegative normalized linear)
sum of projectors corresponding to sum of definite rays with
definite probabilities, which may be chosen to be mutually orthogonal.

In forming superposition of two vectors, their relative phases are important.
So the rays by themselves are insufficient. They suffice to form mixtures.
The question arises as to how to work with the projectors and yet get a
superposition with definite (relative) phases within the formalism of density
operators, either pure or mixed.

This is accomplished using the method of `purification' of a
nonextremal state. Since there are many possible superpositions,
we must have purification of an impure density to any of those
states. This is accomplished by the use of a suitable fiducial
projector; the choice of this projector determines the phases (or
rather phase differences). We show that there are choices of the
fiducial projector that can give any of the superpositions and
that, even with one such superposition, the fiducial projector can
be chosen amongst a continuous set of projectors.

Having accomplished `purification' we turn to another
characteristic property of quantum system, namely, `quantum
entanglement'. If we have a composite system, say $AB$, composed
of two subsystems $A$ and $B$, the generic states of $AB$ contain
information that is not obtained by considering the states of the
subsystems. These may be refered to as nonlocal correlations
between the subsystems $A$ and $B$, which cannot be attributed to
causal connections. (We have processes which have significant
relationship that cannot be accounted causally; somewhat like the
notion `synchronicity' by Carl Jung~\cite{Sudbook}.) 
Schr\"odinger pointed out this as a characteristic property 
representing a quantum system.

We can have correlations between dynamical variables measured in
subsystems in classical dynamics. In a pure classical state, this
automatically gives pure states with definite values for the
subsystem variables. But in the quantum system the situation is
entirely different. A general pure state of the composite system gives
impure states of the subsystems. In this case, the systems $A$ and
$B$ are `entangled'. For example, the singlet state of two
spin-1/2 (particles) is entangled. Any spin component of the
combined system $AB$ gives a zero expectation value 
(singlet is total spin-0), the
individual particles are completely unpolarized --- any spin
component has zero expectation value. Yet there is a definite
correlation --- if one spin is `down', the other one is `up' (and
vice versa) with respect to any direction.

We recognize that if an entangled pure state is considered as 
a state of two subsystems $A$ and $B$, they could be mixed 
with the same rank but with corresponding eigenvalues
and eigenprojectors. So the problem of recovering the
original pure state from the two mixed (impure) states of the subsystems
involves the restoration of nonlocal phase relations, characteristics of
entangled states. We have evolved a method of restoring the pure state using
an entangled fiducial projector. However, in this case, the restoration can
be done if the subsystems have density matrices of the same rank and same
eigenvalues.

We may also recognize that operations on an entangled pure state may
lead to an unentangled (Kronecker product) state for the composite
system which remains pure when restricted to either subsystem. For
example, if we act on the singlet state with the operator of the
difference of two spins, we can obtain a triplet state which may
be unentangled. The point is that the operator acting as the
difference of the spins is itself `nonlocal' in that it acts on
both subsystems together. Similarly, we can take an unentangled 
pure state like the $\pm 1$ states for the
total spin-1, and an entangled pure state
can be obtained by either acting with some component of the difference
of spins on the singlet state, or by a spin operator antisymmetric
in the two spins on the entangled triplet state. Here again the
operator has the ability to entangle. We shall make use of this
property in generating entanglement. The entanglement and the
relative phases are implicit in the fiducial projector chosen.

\section{Pure state addition}

In [6--8]
the rule to add two pure states
determined by their density operators $\rho_1$ and $\rho_2$, which
are projectors, was formulated. This rule corresponds to
superposition of the state vectors $\psi_1$ and $\psi_2$. This
superposition of vectors is a standard tool to describe the
quantum phenomenon of interference. The interference pattern is
sensitive to relative phase of the two vectors. To describe the
relative phase, a fiducial projector $P_0$ was used. By means of
the fiducial projector, the quantum interference can be described
in terms of operators only without using the state vectors. The
addition law of two orthogonal states reads (see [6--8]
where the addition formula was written in
slightly different form)
\begin{equation}\label{psa3}
\rho=p_1\rho_1+p_2\rho_2+\frac{\left(\rho_1P_0\rho_2
+\mbox{h.c.}\right)
\sqrt{p_1p_2}}{\sqrt{\mbox{Tr}\left(\rho_1P_0\rho_2P_0\right)}}
\end{equation}
where positive probabilities (numbers $p_1$ and $p_2$) satisfy the
normalization relation
\begin{equation}\label{psa4}
p_1+p_2=1.
\end{equation}
One can check that the density operator~(\ref{psa3}) is a
projector, i.e.
\begin{equation}\label{psa5}
\rho^2=\rho\qquad\rho^\dagger=\rho \qquad\mbox{Tr}\,\rho=1
\qquad \rho_i\rho\rho_i=p_i\rho_i\qquad (i=1,2)
\end{equation}
if $\rho_1$, $\rho_2$ and $P_0$ are projectors.

We consider the case $\rho_1\rho_2=0$ and $P_0\rho_1\neq 0$ and
$P_0\rho_2\neq 0$.

The composition law~(\ref{psa3}) can be interpreted as the purification of
the impure density operator
\begin{equation}\label{psa6}
\rho_{\rm im}=p_1\rho_1+p_2\rho_2.
\end{equation}
Nevertheless, the real meaning of the ''purification'' consists of the
statement that for two given orthogonal projectors and given fiducial one
which is not orthogonal to the given projectors, sum (\ref{psa3}) is again
a projector.
Relation~(\ref{psa3}) can be extended to describe the purification procedure
for the impure density operator of the form
\begin{equation}\label{psa7}
\rho_{\rm im}=\sum_kp_k\rho_k
\end{equation}
where projectors are such that $\rho_k\rho_m=0$ $(k\neq m)$ 
and positive probabilities $p_k$ satisfy the normalization condition
\begin{equation}\label{psa8}
\sum_kp_k=1.
\end{equation}
Summation in (\ref{psa7}) and (\ref{psa8}) can be considered as summation
over finite set of indices $k$ or over infinite one for the system with
infinite number of states. The generalization of (\ref{psa3}) provides the
purified density operator
\begin{equation}\label{psa9}
\rho=\sum_{k,j}\sqrt{p_kp_j}\,
\frac{\rho_kP_0\rho_j}{\sqrt{\mbox{Tr}\left(
\rho_kP_0\rho_jP_0\right)}}\,.
\end{equation}
Formula~(\ref{psa3}) can be also extended to take into account
that there is a visibility parameter $\gamma$~\cite{Sud3}, a
characteristic of the interference pattern with $0\leq \gamma\leq 1$.
 Equation~(\ref{psa3}) is generalized to the form
\begin{equation}\label{psa10}
\rho=p_1\rho_1+p_2\rho_2+\gamma\,\frac{\sqrt{p_1p_2}
\left(\rho_1P_0\rho_2+\mbox{h.c.}
\right)}{\sqrt{\mbox{Tr}\left(\rho_1P_0\rho_2P_0\right)}}\,.
\end{equation}

For $\gamma=1$, equation~(\ref{psa10}) reduces to equation~(\ref{psa3}).

For $\gamma=0$, one has the impure state~(\ref{psa6}).

Thus, the visibility parameter $\gamma$ is a characteristic of completeness
of the purification procedure of the density operator  or of degree
 of decoherence of the initial pure superposition state~(\ref{psa3}).
 In the case of (\ref{psa3}) or (\ref{psa10}), we have
 $\rho_k\rho\rho_k=p_k\rho_k$ (no sum on $k$).

 Let us consider now two density operators of quantum states
 which are pure nonorthogonal states.
In this case, one has normalization constant and the purification
formula is
\begin{equation}\label{psa11}
\rho=\left[p_1\rho_1+p_2\rho_2+
\frac{\sqrt{p_1p_2}\left(\rho_1P_0\rho_2+
\mbox{h.c.}\right)}{\sqrt{\mbox{Tr}
\left(\rho_1P_0\rho_2P_0\right)}}\right]
{\cal N}^{-1}.
\end{equation}
The normalization constant ${\cal N}$ reads
\begin{equation}\label{psa12}
{\cal N}=1+\frac{2\sqrt{p_1p_2}\mbox{Re}
\left(\mbox{Tr}\,(\rho_1P_0\rho_2)\right)}
{\sqrt{\mbox{Tr}\,(\rho_1P_0\rho_2P_0)}}\,.
\end{equation}

\section{Impure state addition}

We discuss now the addition rule of two mixed states.

In the case of two impure states $\rho_1$ and $\rho_2$, their sum
can be decomposed in terms of orthogonal projectors $R_n$ (i.e.
satisfying $R_nR_m=\delta_{nm}R_n$):
\begin{equation}\label{psa13}
p_1\rho_1+p_2\rho_2=\sum_n\omega_nR_n\qquad \sum_n\omega_n=1
\qquad \omega_n\geq 0.
\end{equation}
One can consider formula (\ref{psa13}) as giving the result of
mixture of pure states  $R_n$ in (\ref{psa7}). Thus the procedure
of addition of impure states can be fulfilled as follows. First,
one writes the sum of impure states as a convex sum of orthogonal
projectors and then carry out the purification given in 
equation~(\ref{psa9}).

In this case, one gets the pure density operator for the result of
''deformed'' addition of two impure states $\rho_1$ and $\rho_2$ with
probabilities $p_1$ and $p_2$ given by the following expression
which is a generalization of equation~(\ref{psa13}):
\begin{equation}\label{psa18}
p_1\rho_1\oplus p_2\rho_2=\sum_{k\,j}
\sqrt{\omega_k\omega_j}\,
\frac{R_kP_0R_j}{\sqrt{\mbox{Tr}
\left(R_kP_0R_jP_0\right)}}
\end{equation}
where $R_k$ are orthogonal eigenprojectors of the density operator
and $\omega_k$ are nonnegative eigenvalues of the density
operator, i.e.
\begin{equation}\label{psa19}
\left(p_1\rho_1+p_2\rho_2\right)R_k=\omega_kR_k.
\end{equation}
There is no sum on $k$.
The fiducial projector $P_0$ is chosen to satisfy the condition
$P_0R_k\neq 0$.

If one has addition of more than two impure states, i.e. the density operator
of impure state has the form $\sum_{s=1}^Np_s\rho_s$, the result of `deformed'
addition rule has the same form as equation~(\ref{psa18}), namely,
\begin{equation}\label{psa20}
\sum_{s=1}^N\oplus p_s\rho_s=\sum_{k\,j} \sqrt{\omega_k\omega_j}\,
\frac{R_kP_0R_j}{\sqrt{\mbox{Tr} \left(R_kP_0R_jP_0\right)}}
\end{equation}
where the eigenprojectors $R_k$ and nonnegative eigenvalues
$\omega_k$ satisfy the equation
\begin{equation}\label{psa21}
\left(\sum_{s=1}^N p_s\rho_s\right)R_k=\omega_kR_k\qquad 
({\rm no~sum~on~}k).
\end{equation}

Thus the purification procedure which is expressed by the deformed
addition rule denoted by the sign $\oplus$ in the left-hand side
of equations~(\ref{psa18}), (\ref{psa20}) is reduced to obtaining
eigenvectors (eigenprojectors) and eigenvalues of the nonnegative
density operator $\sum_s p_s\rho_s$ and applying ansatz with the
fiducial projector $P_0$ $(P_0R_k\neq 0)$ to construct the
nonlinear expression in the right-hand side of
equation~(\ref{psa18}), (\ref{psa20}). This expression provides
the purified density operator.

In the limit case where the initial density operators $\rho_s$ are
orthogonal projectors, equation~(\ref{psa20}) is reduced to
equation~(\ref{psa9}) with obvious replacement
$\omega_k\rightarrow p_k$. One can point out that only one
projector $P_0$ is sufficient to provide $(N-1)$ independent phase
parameters in the case of addition of $N$ orthogonal pure states.
This projector must have nonzero overlap with all added pure state
projectors.  In the case of addition of $N$ impure states, the
number of independent phase parameters, which are contained in
only one fiducial projector $P_0$, depends on the rank of the
density operator $\sum_sp_s\rho_s$ and is one less than the rank.
For clarity, we point out that the obtained
addition formula for impure density operators assumes the existence 
of the fiducial projector. 
This existence is obvious from a geometrical point of view,
but the explicit finding of this fiducial 
projector for given density operators is a different problem. 

\section{Symbols and their star-product}

In quantum mechanics, observables are described by linear
operators acting on the Hilbert space of states. In order to
consider observables as functions, we review first a general
construction~\cite{OlgaStar} and provide general relations and
properties of a map from operators onto functions having in mind a
map of density operator onto a function like Wigner distribution.
Given a Hilbert space $H$ and an operator $\hat A $ acting on this
space, let us suppose that we have a set of operators $\hat U({\bf
x})$ acting on $H$, where a $n$-dimensional vector ${\bf
x}=(x_1,x_2,\ldots,x_n)$ labels the particular operator in the
set. We construct the $c$-number function $f_{\hat A}({\bf x})$
(we call it the symbol of operator $\hat A$ ) using the definition
\begin{equation}\label{eq.1}
f_{\hat A}({\bf x})=\mbox{Tr}\left[\hat A\hat U({\bf x})\right].
\end{equation}
Let us suppose that there must exist a set of operators $\hat D({\bf x})$
such that
\begin{equation}\label{eq.2}
\hat A= \int f_{\hat A}({\bf x})\hat D({\bf x})~d{\bf x}.
\end{equation}
We will consider relations~(\ref{eq.1})
and~(\ref{eq.2}) as relations
determining the invertible map from the operator $\hat A$ onto
function $f_{\hat A}({\bf x})$.

The product (star-product) of two
functions $f_{\hat A}({\bf x})$ and $f_{\hat B}({\bf x})$
corresponding to two operators $\hat A$ and
$\hat B$ is defined by the relations
\begin{equation}\label{eq.5}
f_{\hat A\hat B}({\bf x})=f_{\hat A}({\bf x})*
f_{\hat B} ({\bf x}):=\mbox{Tr}\left[\hat A\hat B\hat U({\bf x})
\right].
\end{equation}
Since the standard product of operators
on a Hilbert space is an associative product,
formula~(\ref{eq.5})
defines an associative product for the functions
$f_{\hat A}({\bf x})$.

\section{Superposition rule in terms of symbols of density operators}

Using formulas~(\ref{eq.1}) and (\ref{eq.2}), one can write
down a composition rule for two symbols $f_{\hat A}({\bf x})$
and $f_{\hat B}({\bf x})$, which determines the star-product
of these symbols bilinear in the two symbols,
\begin{equation}\label{eq.25}
f_{\hat A}({\bf x})*f_{\hat B}({\bf x})=
\int f_{\hat A}({\bf x}'')f_{\hat B}({\bf x}')
K({\bf x}'',{\bf x}',{\bf x})\,d{\bf x}'\,d{\bf x}''.
\end{equation}
The kernel in the integral of (\ref{eq.25}) is
determined by the trace of product of the basic operators,
which we use to construct the map
\begin{equation}\label{eq.26}
K({\bf x}'',{\bf x}',{\bf x})=
\mbox{Tr}\left[\hat D({\bf x}'')\hat D({\bf x}')
\hat U({\bf x})\right].
\end{equation}

Formula~(\ref{eq.26}) can be extended for the case of
the star-product of $N$ symbols of operators
$\hat A_1,\hat A_2,\ldots,\hat A_N$.
Thus one has
\begin{eqnarray}\label{eq.26'}
f_{\hat A_1}({\bf x})*f_{\hat A_2}({\bf x})*\cdots *
f_{\hat A_N}({\bf x})=\int f_{\hat A_1}({\bf x}_1)
f_{\hat A_2}({\bf x}_2)\cdots f_{\hat A_N}({\bf x}_N)\nonumber\\
\times
K\left({\bf x}_1,{\bf x}_2,\ldots,{\bf x}_N,{\bf x}\right)
\,d{\bf x}_1\,d{\bf x}_2\cdots \,d{\bf x}_N
\end{eqnarray}
where the kernel has the form~\cite{OlgaStar}
\begin{equation}\label{eq.26''}
K\left({\bf x}_1,{\bf x}_2,\ldots,{\bf x}_N,{\bf x}\right)=
\mbox{Tr}\left[\hat D({\bf x}_1)\hat D({\bf x}_2)
\cdots\hat D({\bf x}_N)\hat U({\bf x})\right].
\end{equation}
The trace of an operator $\hat A^{N}$ is determined by
the kernel as follows
\begin{eqnarray}\label{eq.26'''}
\mbox{Tr}\, \hat A^N&=&\int f_{\hat A}({\bf x}_1) f_{\hat A}({\bf
x}_2)\cdots f_{\hat A}({\bf x}_N) \nonumber\\ &&\times
\mbox{Tr}\left[\hat D({\bf x}_1)\hat D({\bf x}_2) \cdots\hat
D({\bf x}_N)\right] \,d{\bf x}_1\,d{\bf x}_2\cdots \,d{\bf x}_N\\
\mbox{Tr}\,(\hat A\hat B)&=&\int f_{\hat A}({\bf x}_1) f_{\hat
B}({\bf x}_2) \mbox{Tr}\,\Big[\hat D({\bf x}_1)\hat D({\bf x}_2)
\Big] \, d{\bf x}_1\,d{\bf x}_2.
\end{eqnarray}
When the operator $\hat A$ is a density  operator of a quantum state,
formula~(\ref{eq.26'''}) for $N=2$ determines the purity parameter of 
the state.

Formulas~(\ref{eq.26'}) and (\ref{eq.26''}) can be used to formulate
the addition law of the density operators of orthogonal pure states
$\rho_i$ as addition law for
their symbols $f_{\rho_i}({\bf x})$. In the case of purification of the sum
$\sum_kp_k\rho_k$ of pure states $\rho_k$ by means of a fiducial projector
$P_0$, one has the following symbols:

$f_\rho({\bf x})$ for purified density operator,

$f_{\rho_k}({\bf x})$ for pure state with density operator $\rho_k$,

$f_0({\bf x})$ for fiducial projector $P_0$.

The formula describing quantum interference in terms of symbols of the
density operators reads
\begin{equation}\label{AA1}
f_\rho({\bf x})=\sum_{j\,k}\sqrt{p_kp_j}\,
\frac{\int f_{\rho_k}({\bf x}_1)f_0({\bf x}_2)
f_{\rho_j}({\bf x}_3)K({\bf x}_1, {\bf x}_2, {\bf x}_3, {\bf x})
\,d{\bf x}_1\,d{\bf x}_2\,d{\bf x}_3}
{\sqrt{\int f_{\rho_k}({\bf x}_1)f_0({\bf x}_2)
f_{\rho_j}({\bf x}_3)f_0({\bf x}_4)
k({\bf x}_1, {\bf x}_2, {\bf x}_3, {\bf x}_4)\,d{\bf x}_1\,d{\bf x}_2
\,d{\bf x}_3\,d{\bf x}_4}}
\end{equation}
where the kernel $K({\bf x}_1, {\bf x}_2, {\bf x}_3, {\bf x})$ is
defined by equation~(\ref{eq.26''}) while the kernel 
$k({\bf x}_1, {\bf x}_2, {\bf x}_3, {\bf x}_4)$, 
which determines trace of the product of
$N$ operators $(N=4)$ in terms of their symbols, reads
\begin{equation}\label{AA2}
k({\bf x}_1, {\bf x}_2,\ldots, {\bf x}_N) =\mbox{Tr}\left[{\hat
D}({\bf x}_1)\,{\hat D}({\bf x}_2)\ldots {\hat D} ({\bf
x}_N)\right].
\end{equation}
In the case of purification of the sum of impure states
(\ref{psa20}), one has analogous formula with the replacement
$\rho_k\rightarrow R_k$ and $p_k\rightarrow \omega_k$.

\section{Weyl symbol addition and interference in terms 
of Wigner--Moyal functions}

In this section, we will consider a known
example of Wigner distribution which is related to
 the Heisenberg--Weyl-group representation.
The interference can be described in terms of Wigner functions using
formula~(\ref{AA1}).
The displacement operator
\begin{equation}\label{s3}
\hat D(\alpha_{\bf x})=\exp\left(\alpha_{\bf x} \hat
a^\dagger-\alpha^*_{\bf x}\hat a\right)
\end{equation}
where
\begin{equation}\label{s4}
\alpha_{\bf x}=x_1+ix_2\qquad
\alpha^*_{\bf x}=x_1-ix_2
\end{equation}
and real numbers $x_1$ and $x_2$ are expressed in terms of 
position and momentum as
\begin{equation}\label{s4'}
x_1=\frac{q}{\sqrt 2}\qquad
x_2=\frac{p}{\sqrt 2}
\end{equation}
determines the basic operators defining the Weyl map.
Thus, one has for the basic operators of the map the following
expressions~\cite{OlgaStar}:
\begin{eqnarray}
\hat U({\bf x})&=&2\hat D(\alpha_{\bf x})
(-1)^{\hat a^\dagger\hat a}
\hat D(-\alpha_{\bf x})\label{s12}\\
\hat D({\bf x})&=&\frac{2}{\pi}\,\hat D(\alpha_{\bf x})
(-1)^{\hat a^\dagger\hat a}
\hat D(-\alpha_{\bf x}).
\label{s13}
\end{eqnarray}
The operator $(-1)^{\hat a^\dagger\hat a}$
is the parity operator $(-1)^{\hat a^\dagger\hat a}=P$,
with the matrix elements given in the position
(or momentum) representation by the formula
\begin{equation}\label{s13'}
\langle x\mid\hat P\mid y\rangle=\delta(x+y).
\end{equation}
The Weyl symbol of density operator $\rho$ is defined by 
(\ref{eq.1}) where we use the operator~(\ref{s12}) and
make the replacement $f\rightarrow W$, which is the state 
Wigner function, and it reads
$$
W_\rho(\alpha)=2\,\mbox{Tr}\,\left[\rho\hat D(\alpha)
(-1)^{\hat a^\dagger\hat a}
\hat D(-\alpha)\right]\qquad(\alpha\equiv\alpha_{\bf x}).
$$
To describe the star-product of Weyl symbols, we introduce a 
generalization of notation (\ref{s4'})
$$
{\bf x}_k=(x_{k_1},x_{k_2})\qquad x_{k_1}=\frac{q_k}{\sqrt 2}
\qquad x_{k_2}=\frac{p_k}{\sqrt 2}\,.
$$
Then 
$$ \alpha_k=\frac{1}{\sqrt 2}\left(q_k+ip_k\right)\qquad
k=1,2,\ldots ,N.
$$ 
The kernel of the star-product of $(N-1)$ Weyl symbols has the
form
\begin{equation}\label{s31}
K\left(\alpha_1,\alpha_2,\ldots,\alpha_N\right)=
\mbox{Tr}\left[\hat U({\bf x}_N)
\Pi_{k=1}^{N-1}\hat D({\bf x}_k)\right].
\end{equation}
The kernel can be rewritten in terms of the complex numbers
$\alpha_i \,\left(i=1,2,\ldots,N\right)$ as \cite{OlgaStar}
\begin{eqnarray}\label{s29'}
K\left(\alpha_1,\alpha_2,\ldots,\alpha_N\right)&=&
\frac{2^{N-1}}{\pi^{N-1}}
\exp\left\{\sum_{j>i}^{N-1}\,\sum_{i=1}^{N-1}
2\left(q^{j-i+2-N}\alpha_i\alpha_j^*+
q^{i-j}\alpha_j\alpha_i^*\right)\right.\nonumber\\
&&\left.+\sum_{i=1}^{N-1} 2\left(q^{1-i}\alpha_i\alpha_N^*+
q^{i+1-N}\alpha_N\alpha_i^*\right)\right\}
\end{eqnarray}
where $q=-1$.

Kernel for trace of the product of four operators reads
\begin{eqnarray}\label{AA3}
&&k(\alpha_1,\alpha_2,\alpha_3,\alpha_4)=\frac{4}{\pi^3}\,\delta^{(2)}
(\alpha_1-\alpha_2+\alpha_3-\alpha_4)\nonumber\\
&&\times\exp\Big\{-2\Big[\Big(\alpha_1\alpha_2^*-\alpha_1\alpha_3^*
+\alpha_1\alpha_4^*+\alpha_2\alpha_3^*
-\alpha_2\alpha_4^*+\alpha_3\alpha_4^*\Big)-\mbox{c.c.}
\Big]\Big\}.
\end{eqnarray}
Having the above kernels we can obtain Weyl symbol of pure density
operator (which we found by means of purification of mixture of 
several states) by
inserting the kernels and the Wigner functions into (\ref{AA1}). 
The explicit result for addition of two Wigner functions was given 
in \cite{Sud1} in a different form.

\section{Symplectic tomograms and superposition principle}

Now we consider the example of tomograms.

According to the general scheme, one can introduce for an operator
$\hat A$ the tomographic symbol $f_{\hat A}({\bf x})$, where ${\bf
x}=(x_1,x_2,x_3)\equiv (X,\mu,\nu)$, which we denote here as
$w_{\hat A}(X,\mu,\nu)$ depending on the position $X$ and the
parameters $\mu$ and $\nu$ of the reference frame~\cite{OlgaStar}
$$w_{\hat A}(X,\mu,\nu)=\mbox{Tr}\left[\hat A
\hat U({\bf x})\right].$$
 The operator $\hat U({\bf x})$ is
given by $$ \hat U({\bf x})\equiv \hat U(X,\mu,\nu)=
\delta\left(X-\mu\hat q-\nu \hat p\right)=
|X|\delta\left(1-\frac{\mu \hat q}{X}-\frac{\nu \hat p}{X}\right)
$$ where $\hat q$ and
$\hat p$ are position and momentum operators.

The inverse transform will be of the form $$\hat A=\int w_{\hat
A}(X,\mu,\nu) \hat D(X,\mu,\nu)\,dX\,d\mu\,d\nu$$ where $$\hat
D({\bf x})\equiv\hat D(X,\mu,\nu)=\frac{1}{2\pi}
\exp\left(iX-i\nu\hat p-i\mu\hat q\right).$$ The kernel defining
the star-product of two tomograms $$K({\bf x}'',{\bf x}',{\bf x})=
\mbox{Tr}\left[\hat D(X'',\mu'',\nu'') \hat D(X',\mu',\nu')\hat
U(X,\mu,\nu)\right]$$ reads~\cite{OlgaStar}
\begin{eqnarray*}
&&K(X_1,\mu_1,\nu_1,X_2,\mu_2,\nu_2,X\mu,\nu)=
\frac{\delta\Big(\mu(\nu_1+\nu_2)-\nu(\mu_1+\mu_2)\Big)}{4\pi^2}
\nonumber\\
&&\times
\exp\left[\frac{i}{2}\left(\left(\nu_1\mu_2-\nu_2\mu_1\right)
+2X_1+2X_2
-\frac{2\left(\nu_1+\nu_2\right)}{\nu}\,X\right)\right].\nonumber
\end{eqnarray*}
The trace of product of four basic operators, which provides kernel
to caclulate the denominator in addition formula of density 
operators (\ref{AA1}) reads
\begin{eqnarray}\label{NA}
&&k({\bf x}_1,{\bf x}_2,\ldots,{\bf x}_N)
=\mbox{Tr}\,\left[\Pi_{k=1}^N\hat D\left(X_k,\mu_k,\nu_k\right)
\right]=(2\pi)^{1-N} \delta\left(\sum_{k=1}^N\mu_k\right)
\delta\left(\sum_{k=1}^N\nu_k\right)\nonumber\\
&&\times\exp\left\{i\left(\sum_{k=1}^NX_k+\frac{1}{2}
\sum_{k<s=1}^N(\nu_k\mu_s-\mu_k\nu_s)\right) \right\}
\qquad N=4.
\end{eqnarray}
Having the above kernels we can obtain the tomogram of pure
density operator inserting the kernels into (\ref{AA1}) and
making there the replacement $f\rightarrow w$. In a
different form the explicit result for addition of tomograms was
done in \cite{Sud1}.

\section{Notion of entanglement}

Another quantum-mechanical property related to superposition 
principle of states in bipartite and multipartite systems is 
entanglement.
Let us have density operator $\rho_{AB}$ of composite system $AB$
which has two subsystems $A$ and $B$. This means that there exist
experimental possibilities to measure the properties of the
subsystem $A$ and of the subsystem $B$. The density operator
$\rho_{AB}$ determines two density operators of the subsystems
$$\rho_A=\mbox{Tr}_B\rho_{AB} \qquad \mbox{and} \qquad
\rho_B=\mbox{Tr}_A\rho_{AB}. $$ Let us consider tensor product of
the two subsystem density operators
$$\rho_{A\times B}=\rho_A\otimes\rho_B.$$  
There is difference of two density operators 
$$ R_{AB}=\rho_{AB}-\rho_A\otimes\rho_B.$$ 
This difference is a characteristic of entanglement. If the system 
is in the state $\rho_{AB}$, which is disentangled, the operator
$R_{AB}=0$. Numerical characteristic of  entanglement is
described by nonzero matrix elements of the operator $R_{AB}$. A
basic independent (invariant)
 characteristic  of the operator $R_{AB}$ is the number
\begin{equation}\label{N5}
e=\mbox{Tr}\,\Big(R_{AB}R^\dagger_{AB}\Big).
\end{equation}
This number can be considered as a measure of entanglement. Since
$R^\dagger_{AB}=R_{AB}$ one has
$$e=\mbox{Tr}\left(R^2_{AB}\right).$$ There are other numerical
characteristics of entanglement like traces of higher powers of
the matrix $R_{AB}$ 
$$e^{(n)}=\mbox{Tr}\left(R_{AB}^{n+1}\right).$$
The state $\rho_{AB}$ is characterized by purity parameter
$$\mu_{AB}=\mbox{Tr}\,\rho^2_{AB}$$ and the state $\rho_{A\times
B}$ has its own purity parameter $$\mu_{A\times B}=\mu_A\mu_B$$
where $$ \mu_A=\mbox{Tr}\,\rho_A^2\qquad\mbox{and}\qquad
\mu_B=\mbox{Tr}\,\rho_B^2.$$ 
Since
\begin{equation}
R_{AB}^2=\rho_{AB}^2 +(\rho_A\otimes\rho_B)^2 -\rho_{AB}\rho_A\otimes
\rho_B-\rho_A\otimes\rho_B\rho_{AB}\label{N11}
\end{equation}
one has
\begin{equation}\label{N11a}
\mbox{Tr}\,R_{AB}^2=\mbox{Tr}\,\rho_{AB}^2+\mbox{Tr}\,
\Big(\rho_A^2\otimes\rho_B^2\Big)-2\,\mbox{Tr}\,
(\rho_{AB}\rho_A\otimes\rho_B).
\end{equation}
Thus we get the measure of entanglement in the form
\begin{equation}\label{N12}
e=\mu_{AB}+\mu_A\mu_B-2\sqrt{\mu_{AB}\mu_A\mu_B}\cos\theta.
\end{equation}
The last term in the right-hand side of (\ref{N12})
is determined as
$$\sqrt{\mu_{AB}\mu_A\mu_B}\cos\theta=
\mbox{Tr}\left(\rho_{AB}\rho_A\times\rho_B\right).
$$
For pure state  $\mu_{AB}=1$,  one has $\mu_A=\mu_B=\mu$
(see Appendix)  and it gives
$e=1+\mu^2-2\mu\cos\theta$.
The introduced angle $\theta$ and parameters
$\mu_A,\mu_B$ and $\mu_{AB}$ can be functionally dependent.
In view of this, finding maxima or minima of the entanglement 
measure needs taking into account this dependence.

In fact, the measure of entanglement (\ref{N5}) is defined using
the notion of distance
between two density operators (see, e.g. \cite{WunscheFort}). 
The connection of distance with measure of entanglement is natural
and it was discussed , e.g., in [31--34].
In the present work, we use the Hilbert--Scmidt distance as measure 
of entanglement but the novelty of suggested measure of entanglement 
is related to the choice of the density operators being compared. 
We use the
distance between the system density operator and the tensor product
of the partial traces over the subsystem degrees of freedom.
This characteristic is intrinsic because it is contained in the state
density operator only.
The geometrical sense of the notion of the entanglement measure
 can be clarified using an analogy 
with distance between the points on Eucledean vectors 
$|{\bf a}-{\bf b}|$,
where real vectors ${\bf a}$ and ${\bf b}$ describe the points.
With this definition, a partially separabble system has a nonzero 
entanglement.

Each matrix can be considered as a complex vector. 
The standard scalar product of any two vectors ${\bf C}$
 and ${\bf D}$ can be always treated as
 $${\bf C}\cdot {\bf D}=\sum_sC_s^*D_s.$$
 If one considers the matrix element of the two matrices $C$ and $D$
 as components of the vectors ${\bf C}$ and ${\bf D}$, one also has
 $${\bf C}\cdot{\bf D}= \mbox{Tr}\,(DC^\dagger).
 $$
  For Hermitian matrices $C=C^\dagger$ and $D=D^\dagger$,
  $${\bf C}\cdot{\bf D}=\mbox{Tr}\,(CD).
  $$
If one considers a density matrix as the vector, purity parameter
plays the role of square of the vector length, so one has for the
purity parameter the inequality
  $$o<\mu\leq 1.$$
Description of the matrix $R_{AB}$ as a vector makes obvious that 
measure of entanglement~(\ref{N5}) coincides with square of the vector
length, which in turn is difference of two other vectors. 
  This means that the length of the vectors under consideration
which correspond to normalized density matrices is less than unity. 
Thus, the geometrical interpretation of measure of entanglement 
  means that angle $\theta$ in (\ref{N12}) is the angle between the
  two vectors. This angle can depend on the length of the vectors
  determining the purity parameters of the system and subsystems.
  The angle parameter is introduced only in order to illustrate 
  the geometrical picture of the entanglement measure under discussion.
  
\section{Entanglement for arbitrary observables}

Usually the notion of entanglement is applied for the density operator.
Mathematically the construction of measure of entanglement $e$ given by
(\ref{N5}) can be extended for arbitrary observable $\hat O$ represented 
in the form of sum of projectors
$$\hat O=\sum_na_n\hat P_n$$
where the $\hat P_n$ are eigenprojectors  and $a_n$ are eigenvalues of
the observable $\hat O$, i.e.
$$\hat O\hat P_n=a_n\hat P_n.$$
For a density operator, the eigenvalues are nonnegative numbers.
For an arbitrary observable, the eigenvalues are real numbers and they
can take negative values. If one has prescribed division of the system
in terms of two subsystems $A$ and $B$, the observable $\hat O$ which
acts in Hilbert space of the system can be treated in the same manner
as density operator in previous section. Thus one can define the reduced
observables
$$\hat O_A=\mbox{Tr}_B\,\hat O\qquad
\hat O_B=\mbox{Tr}_A\,\hat O.
$$
The tensor product of the observables
$$\hat O_{A\times B}=\hat O_A\otimes\hat O_B
$$
acts in the Hilbert space of the system.

The correlations of two subsystems captured by the observable $\hat O$
can be connected with a measure of  entanglement like in the case of
density operator. We define the measure of  entanglement for the
observable $\hat O$ as the number
$$e_0=\mbox{Tr}\,\Big[(\hat O-\hat O_A\otimes\hat O_B)^2\Big]=
\mbox{Tr}\,\Big[(\hat O-\hat O_A\otimes\hat O_B)
(\hat O^\dagger-\hat O_A^\dagger\otimes\hat O_B^\dagger)\Big].
$$
This number gives invariant description of a `distance' between
two Hermitian operators $\hat O$ and $\hat O_A\otimes\hat O_B$
exactly in the same manner as in the case of distance between
two density operators.

Analogously, one can introduce positive parameters
$$\mu_0=\sum_n a_n^2\qquad
\mu_{0A}=\sum_k a_{kA}^2\qquad
\mu_{0B}=\sum_\alpha a_{\alpha B}^2$$
where $a_{kA}$ and $a_{\alpha B}$ are eigenvalues of the Hermitian
matrices $\hat O_A$ and $\hat O_B$, respectively.
So formula~(\ref{N12}) can be extended for arbitrary observable in the
form
$$e_0=\mu_0+\mu_{0A}\mu_{0B}-2\sqrt{\mu_0\mu_{0A}\mu_{0B}}\cos\theta
$$
where we define $\cos\theta$ using the same geometrical interpretation
of scalar product of two vectors
$$\sqrt{\mu_0\mu_{0A}\mu_{0B}}\cos\theta=\mbox{Tr}\left(\hat O
\hat O_A\otimes\hat O_B\right).
$$
Thus we introduced notion of entanglement for other Hermitian observables
than for density operators. Of course, the inequalities for purity
parameters in the case of density operators are not valid for other
observables.

One can make generalization introducing the measure of entanglement
of $k$th order of arbitrary observable  
to multipartite system $AB\ldots C$
using the definition of  measure
\begin{equation}\label{E76}
e_0^{(k)}=\mbox{Tr}\Big[(\hat O-\hat O_A\otimes\hat O_B\otimes
\cdots\otimes\hat O_C)^k\Big]
\qquad \hat O_A=\mbox{Tr}_{B\ldots C}\hat O,\ldots\qquad k=2,3,\ldots N.
\end{equation}
For even $k$, the above parameter is nonnegative number.
The measure can be normalized using the factor $\mu_0^{-1}$.

\section{Example of two qubits}

Let us consider a density matrix with unit trace  for two spins in 
the basis
$\mid\uparrow\rangle$ and $\mid\downarrow\rangle$ for the first spin 
and for the second spin, correspondingly, i.e., 
in the basis in four-dimensional space
$$
\mid\uparrow\uparrow\rangle\qquad
\mid\uparrow\downarrow\rangle\qquad
\mid\downarrow\uparrow\rangle\qquad
\mid\downarrow\downarrow\rangle.
$$
The Hermitian density matrix has a form
\begin{equation}\label{L1}
\rho=\left(\begin{array}{clcl}
\rho_{11}&\rho_{12}&\rho_{13}&\rho_{14}\\
\rho_{21}&\rho_{22}&\rho_{23}&\rho_{24}\\
\rho_{31}&\rho_{32}&\rho_{33}&\rho_{34}\\
\rho_{41}&\rho_{42}&\rho_{43}&\rho_{44}
\end{array}\right).
\end{equation}
The density matrix $\rho_A=\mbox{Tr}_B\rho$ reads
\begin{equation}\label{L2}
\rho_A=\left(\begin{array}{clcl}
\rho_{11}+\rho_{22}&\rho_{13}+\rho_{24}\\
\rho_{31}+\rho_{42}&\rho_{33}+\rho_{44}
\end{array}\right)
\end{equation}
and the density matrix $\rho_B=\mbox{Tr}_A\rho$ reads
\begin{equation}\label{L3}
\rho_B=\left(\begin{array}{clcl}
\rho_{11}+\rho_{33}&\rho_{12}+\rho_{34}\\
\rho_{21}+\rho_{43}&\rho_{22}+\rho_{44}
\end{array}\right).
\end{equation}
The tensor product of two matrices
$\rho_A$ and $\rho_B$ has the form quadratic in matrix
elements of the matrix $\rho$
$$
\rho_A\otimes \rho_B=\left|
\begin{array}{clcl}
(\rho_{11}+\rho_{22})\rho_B&(\rho_{13}+\rho_{24})\rho_B\\
(\rho_{31}+\rho_{42})\rho_B&(\rho_{33}+\rho_{44})\rho_B
\end{array}\right|.
$$
The purity parameter of the two-spin state (\ref{L1})
equals
\begin{equation}\label{L6}
\mu=\sum_{i,k=1}^4|\rho_{ik}|^2.
\end{equation}
The purity parameter of the states (\ref{L2}) and (\ref{L3})
read
\begin{equation}\label{L7}
\mu_A=|\rho_{11}+\rho_{22}|^2+|\rho_{13}+\rho_{24}|^2
+|\rho_{31}+\rho_{41}|^2+|\rho_{33}+\rho_{44}|^2
\end{equation}
and
\begin{equation}\label{L8}
\mu_B=|\rho_{11}+\rho_{33}|^2+|\rho_{12}+\rho_{34}|^2
+|\rho_{21}+\rho_{43}|^2+|\rho_{22}+\rho_{44}|^2.
\end{equation}
One can calculate the trace 
defining the angle between two vectors corresponding to the
density operators 
in the form
\begin{eqnarray}\label{L9}
&&\mbox{Tr}\,\Big(\rho\rho_A\otimes\rho_B\Big)=\nonumber\\
&&(\rho_{11}+\rho_{22})\Big[\rho_{11}(\rho_{11}
+\rho_{33})+\rho_{12}(\rho_{21}+\rho_{43})+
\rho_{21}(\rho_{12}+\rho_{34})
+\rho_{22}(\rho_{22}+\rho_{44})\Big]\nonumber\\
&&+(\rho_{31}+\rho_{42})\Big[\rho_{13}(\rho_{11}
+\rho_{33})+\rho_{14}(\rho_{21}+\rho_{43})+
\rho_{23}(\rho_{12}+\rho_{34})
+\rho_{24}(\rho_{22}+\rho_{44})\Big]\nonumber\\
&&+(\rho_{13}+\rho_{24})\Big[\rho_{31}(\rho_{11}
+\rho_{33})+\rho_{32}(\rho_{21}+\rho_{43})
+\rho_{41}(\rho_{12}+\rho_{34})
+\rho_{42}(\rho_{22}+\rho_{44})\Big]\nonumber\\
&&+(\rho_{33}+\rho_{44})\Big[\rho_{33}(\rho_{11}
+\rho_{33})+\rho_{34}(\rho_{21}+\rho_{43})
+\rho_{43}(\rho_{12}+\rho_{34})
+\rho_{44}(\rho_{22}+\rho_{44})\Big].\nonumber\\
&&
\end{eqnarray}
Having expressions (\ref{N12}), (\ref{L6})--(\ref{L9})
one can calculate measure of entanglement for arbitrary density 
matrix of two spins.
For example, in the case of pure state
\begin{equation}\label{L10}
\rho=\frac{1}{2}\left(\begin{array}{clcl}
0&0&0&0\\
0&1&1&0\\
0&1&1&0\\
0&0&0&0\end{array}\right)
\end{equation}
one has 
\begin{equation}\label{L11}
\mu=1\qquad\mu_A=\frac {1}{2}\qquad\mu_B=\frac{1}{2}\,.
\end{equation}
This provides the maximum entanglement of the state~(\ref{L10}),
i.e.
\begin{equation}\label{L12}
e=\frac{3}{4}\,.
\end{equation}
For more general matrix of pure state of the form
\begin{equation}\label{L13}
\rho=\left(\begin{array}{clcl}
0&0&0&0\\
0&c^2&sc&0\\
0&sc&s^2&0\\
0&0&0&0\end{array}\right)
\end{equation}
where $c\equiv\cos\varphi$ and $s\equiv\sin\varphi$, one gets
\begin{equation}\label{L14}
e_\varphi=\frac{1}{2}\sin^22\varphi
\left[1+\frac{\sin^22\varphi}{2}
\right].
\end{equation}
The angle in (\ref{N12}) is determined by the angle $\varphi$ of
(\ref{L14}). For $\varphi=45^o$, one has $\theta=60^o$.

\section{Distance and entanglement of Gaussian squeezed states}

One can use the developed approach to study entanglement of two-mode 
squeezed Gaussian states.
Wigner function of generic squeezed and correlated state
$\rho$ in $n$ dimensions has the form $(\hbar=1)$~\cite{Vol183}
\begin{equation}\label{S1}
W({\bf Q})=
N\exp\left[-\frac{1}{2}\left(\underline{\bf Q}\right)
\Sigma^{-1}\left(\underline{\bf Q}\right)
\right]
\qquad N=(\det\Sigma)^{-1/2}
\end{equation}
where ${\bf Q}=(p_1,p_2,\ldots,p_n,q_1,q_2,\ldots,q_n)$ and
\begin{equation}\label{S2}
\underline{\bf Q}={\bf Q}-\langle{\bf Q}\rangle
\end{equation}
with $\langle{\bf Q}\rangle$ being the parameters describing means of the
quadature components.
The 2$n$$\times$2$n$ matrix $\Sigma$ describes variances and covariances 
of the quadrature components.
We present the dispersion matrix in the form
\begin{equation}\label{S9}
\Sigma=\left(\begin{array}{clcl}
\Sigma_1&\Sigma_{12}\\
\Sigma_{12}^t&\Sigma_2\end{array}\right)
\qquad
\Sigma^{-1}=\left(\begin{array}{clcl}
A&B\\
B^t&C\end{array}\right).
\end{equation}
We consider two subsystems with dimensions $n_1$ and $n_2$, with $n_1+n_2=n$.
Let us suppose that the system state has the parameters
$\langle{\bf Q}\rangle=0$ (squeezed vacuum in the case of pure states). 
The normalization constant 
$N$ in (\ref{S1}) is determined by the matrix $\Sigma$ due to the condition
\begin{equation}\label{S3}
\int W({\bf Q})\frac{d{\bf Q}}{(2\pi)^n}=1.
\end{equation} 
The purity parameter of the Gaussian state of the system
equals~\cite{Vol183}
\begin{equation}\label{S3'}
\mu=\mbox{Tr}\,\rho^2=\int W^2({\bf Q})\frac{d{\bf Q}}{(2\pi)}^n=
2^{-n}(\det\Sigma)^{-1/2}.
\end{equation}
Integrals~(\ref{S3}) and (\ref{S3'}) can be calculated using the formula
for $n$-dimensional Gaussian integral
\begin{equation}\label{S4}
\int e^{-{\bf x}a{\bf x}+{\bf {bx}}}d{\bf x}=
\frac{\pi^{n/2}}{\sqrt{\det a}}\exp\left(\frac{1}{4}{\bf b}
a^{-1}{\bf b}\right).
\end{equation}
The Wigner function of the subsystem state 1 $\rho_1$, which is denoted
 as $W_1({\bf Q}_1)$, is given by the relation
\begin{equation}\label{S5}
W_1({\bf Q}_1)=
\int W({\bf Q})\frac{d{\bf Q}_2}{(2\pi)^{n_2}}
\end{equation} 
and the Wigner function of the subsystem state 2 $\rho_2$
is given by analogous relation
\begin{equation}\label{S6}
W_2({\bf Q}_2)=
\int W({\bf Q})\frac{d{\bf Q}_1}{(2\pi)^{n_1}}.
\end{equation} 
Both integrals are Gaussian ones. Due to this, one has
\begin{eqnarray}
W_1({\bf Q}_1)=
N_1\exp\left[-\frac{1}{2}\Big({\bf Q}_1\sigma_1^{-1}{\bf Q}_1\Big)\right]
&\qquad & N_1=(\det\sigma_1)^{-1/2}\label{S7}\\
W_2({\bf Q}_2)=
N_2\exp\left[-\frac{1}{2}\Big({\bf Q}_2\sigma_2^{-1}{\bf Q}_2\Big)\right]
&\qquad & N_2=(\det\sigma_2)^{-1/2}\label{S7'}
\end{eqnarray}
where 
$$
\sigma_1^{-1}=A-BC^{-1}B^t\qquad\sigma_2^{-1}=C-B^tA^{-1}B.
$$
The purity parameters of the states of subsystem read
\begin{equation}\label{S8'}
 \mu_1=2^{-n_1}(\det\sigma_1)^{-1/2}\qquad
  \mu_2=2^{-n_2}(\det\sigma_2)^{-1/2}.
  \end{equation}
In the case $\det\Sigma=(1/4)^n$, $\mu_1=\mu_2$.

The normalization constants $N_1$ and $N_2$ are functions of initial 
dispersion matrix $\Sigma$. The Wigner function of the state 
$\rho_1\otimes\rho_2$ has the product form 
\begin{equation}\label{S10}
W_{12}({\bf Q})=W_1({\bf Q}_1)W_2({\bf Q}_2).
\end{equation} 
This form is also Gaussian one
\begin{equation}\label{S11}
W_{12}({\bf Q})=
N_{12}\exp\left[-\frac{1}{2}\Big({\bf Q}\sigma^{-1}{\bf Q}\Big)\right]
\qquad N_{12}=N_1N_2\qquad
\sigma=\left(\begin{array}{lrlc}
\sigma_1&0\\
0&\sigma_2\end{array}\right).
\end{equation}
To calculate introduced measure of entanglement, one has to calculate
the fidelity $t=\mbox{Tr}\Big(\rho\rho_1\otimes\rho_2\Big)$, which is
expressed in terms of Wigner functions by the integral
\begin{equation}\label{S13}
t=\int \frac{d{\bf Q}}{(2\pi)^{n}}
W({\bf Q})W_{12}({\bf Q}).
\end{equation} 
The integral is Gaussian again with the dispersion parameters 
$(\sigma +\Sigma)$. So one has
$$t=\Big(\det\,(\Sigma +\sigma)\Big)^{-1/2}.$$
This trace determines the term with $\cos\theta$ in the expression 
for entanglement of the squeezed Gaussian state.

Thus measure of entanglement of the squeezed Gaussian state reads
$$e_{\rm G}=2^{-n}(\det\Sigma)^{-1/2}+2^{-n}(\det\sigma)^{-1/2}
-2\Big(\det(\Sigma+\sigma)\Big)^{-1/2}.$$
It is determined by the quadrature dispersion matrix of the composite
system, which is characteristic of Gaussians. 
If the pure state is squeezed but not 
correlated~\cite{Kurm,SudBham},
entanglement is absent. For entanglement, one needs the correlation of
quadratures in the initial pure states.\footnote{It is worthy noting 
that another measure of entanglement based on the cross covariances 
of quadrature components of entangled modes was introduced
in \cite{DodCastroJRLR}.}

\section{Purification of separable density matrix}

In this section, we consider the procedure of purification of mixed
density matrix. Density matrix of a composite system is said to be 
simply separable if it has the form
\begin{equation}\label{ps1}
\rho_{AB}=\rho_A\otimes\rho_B
\end{equation}
where $\rho_A=\mbox{Tr}_B\rho_{AB}$ and  
$\rho_B=\mbox{Tr}_A\rho_{AB}$. 
Such matrix can be pure if and only if $\rho_A$ and $\rho_B$   
are pure and hence 
$\rho_{AB}$, $\rho_A$ and $\rho_B$ are projectors in the appropriate 
spaces. Such a pure state of the composite system is not entangled. 
More generally, a density matrix $\rho_{AB}$ is said to be separable 
if
\begin{equation}\label{ps2}
\rho_{AB}=\sum_n p_n\rho_{nA}\otimes\rho_{nB}\qquad \sum_np_n=1
\qquad p_n\geq 0.
\end{equation}
In this case, 
$$\rho_{A}=\sum_n p_n\rho_{nA}\qquad \mbox{and}
\qquad \rho_{B}=\sum_n p_n\rho_{nB}.$$
Clearly if $n\geq 2$, the density matrix $\rho_{AB}$ is not pure.
For $n=1$ the matrix $\rho_{AB}$ is not pure unless $\rho_A$  and 
$\rho_B$ are one-dimensional projectors. 
Since $\rho_{nA}$ and $\rho_{nB}$ are density matrices, 
they are convex linear sums of projectors
\begin{equation}\label{ps3}
\rho_{nA}=\sum_j p_{nj}\Pi_{nj}^A
\qquad\mbox{and}\qquad
\rho_{nB}=\sum_k q_{nk}\Pi_{nk}^B
\end{equation}
where $p_{nj}$ and $q_{nk}$ are nonnegative numbers
$\sum_j p_{nj}=\sum_k q_{nk}=1$.
Thus $\rho_{AB}$ is the convex sum of projectors
\begin{equation}\label{ps4}
\rho_{AB}=\sum_{n,j,k} p_{nj}
q_{nk}u_{nj}v_{nk}u_{nj}^\dagger v_{nk}^\dagger.
\end{equation}
Note, these projectors are not all mutually orthogonal and $\rho_{AB}$
is a mixture with weight $p_{nj}q_{nk}$ for all $n,j,k$.
The density matrices
\begin{equation}\label{ps5}
\rho_{AB}(n,j,k)=u_{nj}v_{nk}u_{nj}^\dagger v_{nk}^\dagger
\end{equation}
(there is no sum) are not mutually orthogonal pure state projectors, 
except for $n=1$. So, once the eigenvectors (eigenrays) are given, we
could construct the density matrix of $AB$ as a convex sum of projectors.
For $n=1$, what does $\rho_{AB}$ contain that is not contained in 
$\rho_A\otimes\rho_B$? 
In this case, it is a set of phase diferences between  various
eigenvectors that go to make up  $\rho_{AB}$. There are 
$\mbox{rank}(\rho_A)\times\mbox{rank}(\rho_B)$ phases 
and hence one less phase differences.
All this is very similar to the purification of mixed states. We write
\begin{equation}\label{ps6}
\rho_{A}=\sum_j p_{j}\Pi_{j}^A
\qquad\rho_{B}=\sum_k q_{k}\Pi_{k}^B
\end{equation}
and hence by hypothesis
\begin{equation}\label{ps7}
\rho_{AB}=\rho_A\otimes\rho_B=\sum_{jk} p_{j}q_k\Pi_{j}^A
\otimes\Pi_{k}^B.
\end{equation}
So, for purification, we adopt the ansatz given earlier (we omit the Kronecker
product symbol)
\begin{equation}\label{ps8}
\widetilde\rho_{AB}=\sum_{jkj'k'}\Big(p_{j}q_kp_{j'}q_{k'}\Big)^{1/2}
\frac{\Pi_{j}^A\Pi_{k}^B\Pi^{AB}\Pi_{j'}^A\Pi_{k'}^B}{
\sqrt{\mbox{Tr}\,\left(\Pi_{j}^A\Pi_{k}^B\Pi^{AB}\Pi_{j'}^A
\Pi_{k'}^B\Pi^{AB}\right)}}\,.
\end{equation}
This $\widetilde\rho_{AB}$ is a pure matrix with probability weights 
$p_jq_k$  for the projectors $\Pi_j^A\Pi_k^B$  and the suitable 
phase differences which number
$\mbox{rank}(\rho_A)\times\mbox{rank}(\rho_B)-1$.
It is essential to choose $\Pi_{AB}$ such that
 $\mbox{Tr}\,\left(\Pi^{AB}\Pi_{j}^A\Pi_{k}^B\right)\neq 0$
 for all $j,k$.
 But the density matrix so constructed will \underline{not} 
 lead to $\rho_A$ and $\rho_B$ as partial traces.
 
 For a more general case of $n\geq 2$, we have the problem of 
 rediagonalizing
 \begin{equation}\label{ps9}
\rho_{A}=\sum_{n,j} p_{nj}\Pi_{nj}^A\qquad
\rho_{B}=\sum_{n,k} q_{nk}\Pi_{nk}^B.
\end{equation}
Once this is done, we proceed as in the simply separate case (n=1)
discussed above.

\section{Entanglement and straddling of fiducial projectors}

The fiducial projector in the purification protocol generates the relative
phases between the (two or more) density matrices for pure states which have 
been `superposed'. Such an operator is a `phase correlator'. The Hermitian
fiducial projector and the projectors which are being superposed are all
Hermitian, yet relative phases are introduced. 
Given two or more Hermitian matrices one can generate the Bargman phase from 
their product. For this to occure, the operators cannot all commute
(phase (ABC) $\neq$ 0). Simple examples may be provided by the Pauli matrices 
(or Dirac matrices), in our case, the phase difference between amplitudes 
is generated by the overlap of the fiducial matrix with the respective density 
matrices. Since we do a normalization in (12)  only the phase of the overlap 
survives. There is a source of the phase interference introduced in our
composition law (12). The question naturally arises --- can we
choose the fiducial projector $P_0$ so as to produce any set of phase 
differences? The answer is affirmative but not unique.

The fiducial projector must
straddle the pure states which are added, that is, it must have nonzero overlap
with each of them. (It may or may not have overlap with other states.)
Following up on this notion we find that the fiducial projector which restores
a fully entangled pure state of a composite system straddles the 
eigenprojectors 
of the  individual rays which are direct product of pure density matrices.
(This is for separable systems, otherwise we get some direct product pure states 
and some fully entangled pure states.) This straddling implies that automatically 
the fiducial projector is a `nonlocal' operator acting coherently on the subsystems.
Since it is also a projector, it follows that this projector is a fully entangled 
pure state. Only such an entangled projector can restore full entanglement.
 Fully entangled operators can be multiplied by each others or added together
 to obtain fully entangled operators, but they will not be projectors of rank one.
 This entanglement (and phase coherence) can be inherent in operators as well
 as in states.

\section{Conclusions}

We presented an intrinsic approach to different quantum phenomena which are 
entanglement and interference. The approach is intrinsic because it points out 
the unique basis for both phenomena which is superposition principle of quantum 
states. But to use this superposition principle in generic case of mixed states, 
one needs the addition formula for density operators. The discussed measure of 
entanglement is intrinsically connected with the given state of a composite system 
because it is determined completely by partial traces of the state density 
operator and by the deviation of the density operator from the tensor product
of the partial traces. Thus, because the entanglement is the property related
to the state superpositions (expressed in terms of new addition rule of 
density operators with using a fiducial projector) 
the fiducial projector becomes a useful tool for 
treating both phenomena --- interference and entanglement.

To conclude, we point out new results of the paper.

The nonlinear addition rule for impure density matrices, which
results in pure density matrix given in (\ref{psa18}) and (\ref{psa20}), 
is a new purification procedure.
The addition rule formula (\ref{AA1}) for symbols of 
density operators of any kind (including Wigner distribution, tomograms, 
etc.) is another new result of our consideration.

The notion of measure of entanglement of arbitrary order for bipartite and 
multipartite systems for an arbitrary observable given by (\ref{E76}) 
is a new aspect of entanglement suggested in our study.
As a partial case, the measure introduced contains the description 
of measure of entanglement of density operator for bipartite system 
given in (\ref{N12}). 
The measure of entanglement is related directly to intrinsic properties 
of density operator of a composite system.

\section*{Acknowledgments}

V~I~M and E~C~G~S thank Dipartimento di
Scienze Fisiche, Universit\'a ``Federico~II''
di Napoli and Istitito Nazionale di Fisica Nucleare,
Sezione di Napoli for kind hospitality.
V~I~M is grateful to the Russian
Foundation for Basic Research for partial support
under Project~No.~01-02-17745.

\section*{Appendix~1.~Partial density matrices for pure state of
composite system}

Let us consider the pure state of a composite system which has 
two subsystems $A$ and $B$. 
The pure state is described by a vector of the form
\begin{equation}\label{E1}
\psi=\sum_{i=1}^N\sum_{\alpha=1}^MC_{i\alpha}\varphi_i\chi_\alpha
\end{equation}
where  $N$ is dimension of the subsystem $A$, $M$ is dimension of the 
subsystem $B$, and the orthogonal vectors $\varphi_i$ $(i=1\ldots N)$
and $\chi_\alpha$ $(\alpha=1\ldots M)$ form basis in Hilbert spaces 
of the subsystem states.

The density operator of the pure state which corresponds to the 
decomposition (\ref{E1}) of the state vector has the form  
\begin{equation}\label{E2}
\rho_{AB}=
\psi\psi^\dagger=\sum_{i,j=1}^N~\sum_{\alpha,\beta=1}^M
C_{i\alpha}C^*_{j\beta}\,\varphi_i\varphi_j^\dagger
\chi_\alpha\chi_\beta^\dagger.
\end{equation}
The density matrix of the $A$-subsystem state in the chosen basis 
has the matrix elements expressed in terms of decomposition coefficients
\begin{equation}\label{E3}
(\rho_{A})_{ij}=
\sum_{\alpha=1}^M
C_{i\alpha}C^*_{j\alpha}.
\end{equation}
The density matrix of the $B$-subsystem state in the chosen basis 
has the matrix elements
\begin{equation}\label{E4}
(\rho_{B})_{\alpha\beta}=
\sum_{i=1}^N
C_{i\alpha}C^*_{i\beta}.
\end{equation}
Both density matrices $\rho_A$ and $\rho_B$ are nonnegative 
Hermitian matrices and Tr$\,\rho_A=$Tr$\,\rho_B=1$. 
Let us calculate parameters
\begin{equation}\label{E5}
\mu_n^{(A)}=\mbox{Tr}\left(\rho_A\right)^n\qquad
\mu_n^{(B)}=\mbox{Tr}\left(\rho_B\right)^n
\end{equation}
for arbitrary integer $n$.

One can easily see that
\begin{equation}\label{E6}
\mu_n^{(A)}=\mu_n^{(B)}.
\end{equation}
In fact, 
\begin{equation}\label{E7}
\mu_n^{(A)}=
\sum_{i_1,i_2,\ldots i_M=1}^N
~\sum_{\alpha_1,\alpha_2,\ldots \alpha_n=1}^M
C_{i_1\alpha_1}C^*_{i_2\alpha_1}C_{i_2\alpha_2}C^*_{i_3\alpha_2}
\cdots C_{i_{n-1}\alpha_{n-1}}C^*_{i_n\alpha_{n-1}}
C_{i_n\alpha_n}C^*_{i_1\alpha_n}
\end{equation}
and
\begin{equation}\label{E8}
\mu_n^{(B)}=
\sum_{i_1,i_2,\ldots i_n=1}^N
~\sum_{\alpha_1,\alpha_2,\ldots \alpha_n=1}^M
C_{i_1\alpha_1}C^*_{i_1\alpha_2}C_{i_2\alpha_2}C^*_{i_2\alpha_3}
\cdots C_{i_{n-1}\alpha_{n-1}}C^*_{i_{n-1}\alpha_{n}}
C_{i_n\alpha_n}C^*_{i_n\alpha_1}.
\end{equation}
The terms without star are the same in both expressions (\ref{E7}) and
(\ref{E8}). These terms are invariant with respect to arbitrary 
permutations
$$1,2,\ldots, n\rightarrow s_1,s_2,\ldots, s_n.$$
The terms with star look differently in (\ref{E7}) and (\ref{E8}), 
but since both sums (\ref{E7}) and (\ref{E8}) are invariant with
respect to arbitrary permutations,
let us make the particular permutation
$$1,2,\ldots, n-1, n\rightarrow 2, 3,\ldots n,1$$
in sum~(\ref{E8}).
The terms without star are invariant and the terms with star in (\ref{E8})
after permutation coincide with the terms with star in (\ref{E7}). 
This proves equality (\ref{E6}) which means that eigenvalues and 
rank of the matrices $\rho_A$ and $\rho_B$ are the same.
 It is clear that the proof can be extended to multipartite composite
 system $AB\ldots C$. Thus we get the following result.
 Given a pure state of a multipartite quantum system  $\rho_{AB\ldots C}$.
 Then the eigenvalues and ranks of the density matrices  $\rho_A$,
 $\rho_B,\ldots \rho_C$ are equal.


\end{document}